\documentclass[fleqn,10pt]{wlscirep}
\usepackage[utf8]{inputenc}
\usepackage[T1]{fontenc}
\usepackage{lineno}
\usepackage{jabbrv}

\title{Helioseismic Evidence That the Solar Dynamo Originates near the Tachocline}

\author[1,*]{Krishnendu Mandal}
\author[1,2]{Alexander G. Kosovichev}

\affil[1]{New Jersey Institute of Technology, Newark, NJ 07102, USA}
\affil[2]{NASA Ames Research Center, Moffett Field, CA 94035, USA}

\affil[*]{krishnendu.mandal@njit.edu}

\begin{abstract}
 The exact location of the solar dynamo remains uncertain—whether it operates primarily in the near-surface shear layer, throughout the entire convection zone, or near the tachocline - a region of sharp transition in the solar rotation, located at the base of the convection zone, approximately 200,000 km beneath the surface. Various studies have supported each of these possibilities. Notably, the solar magnetic `butterfly' diagram and the pattern of zonal flows (`torsional oscillations') exhibit strikingly similar characteristics, suggesting a link between magnetic field evolution and solar flows. Since magnetic fields cannot be measured directly in the deep solar interior, torsional oscillations and rotation gradients are employed as diagnostic proxies. Our analysis reveals that the gradient of rotation displays `butterfly'–like behavior near the tachocline, which is similar to the magnetic butterfly diagram at the surface. This result supports the idea that the solar dynamo has a deep-seated origin, likely operating either near the tachocline or throughout the convection zone, thereby disfavoring the recent scenario of a shallow, near-surface dynamo. This finding may also have important implications for understanding how stellar dynamos operate in general. 
\end{abstract}

\begin{document}
\flushbottom
\maketitle

\thispagestyle{empty}

\section*{Introduction}
Sunspots remain the most widely recognized manifestation of the solar magnetic activity cycle. Galileo and other early observers were the first to systematically study their occurrence, noting that they typically appear below $30^\circ$ latitude. Nearly two centuries later, Schwabe \cite{schwabe_1843} discovered that the number of sunspots varies with an approximate 11-year periodicity, marking the first detection of the solar cycle. Carrington \cite{carrington_1858} subsequently showed that sunspots first emerge at mid-latitudes and then gradually migrate toward the equator, giving rise to the characteristic “butterfly diagram.” Hale \cite{hale_1908} later demonstrated that sunspots are caused by strong magnetic fields, and further established that they exhibit a bipolar magnetic structure \cite{hale_1919}, with the leading spot polarity being opposite in the two hemispheres. A systematic tilt of the leading and trailing polarities with respect to the east–west direction, now known as Joy’s law, was also observed.

In the latter half of the 20th century, considerable effort was devoted to explaining these observations. A purely magnetic field would dissipate over time through Joule heating, implying that a continuous flow is required to sustain the Sun’s magnetism. This realization spurred intense interest in dynamo theory, which seeks to explain the generation and maintenance of the solar magnetic field through magneto-hydrodynamic processes. In the latter part of the previous century, it was established that the Sun exhibits large-scale flows, most notably differential rotation in the west–east direction and meridional circulation in the north–south direction. Helioseismology provides a means to probe the structure and dynamics of the solar interior. In particular, the splitting of acoustic-mode frequencies reveals large-scale flows such as differential rotation \cite{schou98,antia_1998}, with the equator rotating faster than the poles—a phenomenon that extends throughout much of the solar interior. Structurally, the solar interior is divided into three main regions: the outer convection zone, which occupies the outermost $~30\%$ of the Sun's radius; the radiative zone beneath it; and the central core, where nuclear fusion generates the energy that powers and sustains the Sun’s luminosity. Separating the convection and radiative zones is a thin transitional layer known as the tachocline. In this region, differential rotation shifts abruptly toward nearly solid-body rotation in the radiative interior. The strong shear associated with this transition is widely believed to play a crucial role in driving the solar dynamo. There is also another shear layer just close to the surface known as the near-surface shear layer (NSSL).

The importance of these large-scale flows on the solar dynamo has been extensively studied and has led to the origin of different dynamo models. A central question in solar physics is where the solar dynamo operates—whether in the near-surface shear layer \cite{vasil2024}, throughout the entire convection zone, or primarily in the tachocline \cite{arnab_1990,dikpati09,gustavo16}. Numerous numerical simulations have investigated these processes, and many reproduce key features of the solar magnetic cycle, such as the butterfly diagram. However, the ability of different models to generate similar observational signatures highlights a degeneracy in current dynamo theories. The prevailing paradigm for the solar dynamo is the 22-year Babcock–Leighton model \cite{leighton69}, which describes the cyclic conversion between toroidal and poloidal magnetic fields. During solar maximum, the poloidal field reaches a minimum, while it peaks near solar minimum. Differential rotation stretches the poloidal field, regenerating the toroidal component and sustaining the cycle \cite{parker_55_dynamo,dikpati_1999,brandenburg_2005}. Meridional circulation is thought to play a crucial role by advecting the solar magnetic field at the surface: as sunspots decay, their remnant magnetic flux is carried poleward by this flow. Since the trailing polarity of active regions typically has the opposite sign to the polar field, this advection process facilitates magnetic field reversal. Some models additionally require an equatorward meridional circulation near the tachocline to transport the toroidal magnetic field equatorward as the solar cycle progresses, thereby producing the butterfly diagram. While the poleward meridional flow at the surface is well established, the structure of meridional circulation in the solar interior remains debated. Helioseismic inferences suggest models ranging from a single-cell circulation pattern \cite{rajaguru15,mandal18,gizon20_mer} to more complex multi-cellular structures, such as a double-cell configuration \cite{zhao13,chen17}. Though meridional flow acts as an advection of the solar magnetic field, shear in the solar differential rotation is thought to be essential for generating and sustaining the dynamo \cite{dikpati_1999,brandenburg_2005, pipin2011,karak2016}. This shear is concentrated in two key regions: the near-surface shear layer (NSSL) and the tachocline. The alternative dynamo model \cite{parker_55_dynamo} suggests that the poloidal magnetic field is generated by cyclonic convection at the bottom of the convection zone and stored in the tachocline.  However, there has been no observational evidence that the tachocline is involved in the solar dynamo. 

A dynamo model must explain how solar interior flows evolve during the solar cycle, as both flows and magnetic fields are coupled through the magnetohydrodynamic (MHD) equations \cite{charbonneau_2020}. It should account for the reversal of the polar magnetic field every 11 years, and Hale’s polarity law \cite{hale_1919}, whereby the leading sunspot polarity is opposite in the two hemispheres, the origin of Joy’s law (the systematic tilt of leading and trailing polarity with respect to east-west direction). Additionally, it must explain the equatorward migration of magnetic fields that is observed as the cycle progresses. Since we are not yet there to probe the magnetic field in the solar interior directly. We need to rely on indirect ways to understand how the solar magnetic field evolves in the solar interior. An important indirect probe is the evolution of solar differential rotation with the solar cycle, since large-scale flows, such as differential rotation and meridional circulation, play a key role in the generation and transport of magnetic fields. The magnetic field, in turn, modifies these flows through Lorentz-force back-reaction, reflecting the nonlinear coupling inherent in the solar dynamo. Surface measurements reveal a torsional oscillation pattern superimposed on the mean differential rotation, characterized by alternating bands of faster and slower rotation relative to the mean \cite{howard80,kosovichev_1997,howe_2000}. This pattern has two distinct branches, both emerging near mid-latitudes just before the onset of a solar cycle—one branch migrates toward the equator, while the other moves poleward. There is a strong correlation between solar torsional oscillations observed on the surface and the magnetic butterfly diagram \cite{howe_2009_lrsp}. Previous helioseismic measurements showed that the torsional oscillations extend into the deep convection zone \cite{sasha2019,mandal24_dw,howe18,vorontsov02}.  However, it was not convincing evidence of the torsional oscillations in the tachocline. It is due to poor resolution as we go deeper, uncertainty increases as fewer modes can propagate in deeper layers. Therefore, while the Near-Surface Shear Layer (NSSL) can be investigated with good accuracy \cite{antia2022,cristina2024,komm2023,Rozelot2025,mandal_25}, conducting a similar analysis in the tachocline proves more challenging due to a poor signal-to-noise ratio. We overcome this limitation by analyzing longer time series of data, such as $4\times 72-$days datasets currently available \cite{sylvain23}, instead of $1\times 72-$days datasets traditionally used in the  studies of solar differential rotation. If a rotational shear in the tachocline is responsible for all the conversion and storage of toroidal field during the solar cycle, as found in the study \cite{gustavo16}, its properties must vary with the solar cycle \cite{basu19,basu_2024}, These properties include   changes in the differential rotation between the convection zone and the radiative interior.  Motivated by these, we aim to investigate how the solar zonal flow (torsional oscillation) varies with the solar cycle throughout the convection zone. Specifically, we examine how the gradients of the angular velocity, such as radial gradient $\partial \Omega/\partial r$, and latitudinal gradient, $\partial\Omega/\partial \theta$  and dimensionless radial gradient, $\nabla_{r}(\Omega)=r\partial \Omega/(\Omega \partial r)=\partial \log \Omega/\partial \log r$ evolve over the solar cycle. The gradient of rotation is particularly sensitive in regions where the rotation rate changes rapidly with radius—most notably in the near-surface shear layer (NSSL) and the tachocline. 

\section*{Results}
Solar acoustic modes are characterized by three quantum numbers $(n,\ell,m)$, where $n$ denotes the number of nodes of the eigenfunction in the radial direction, $m$ represents the azimuthal order, and $\ell-\vert m\vert$ gives the number of latitudinal nodes. In the absence of rotation, the mode frequencies are independent of $m$. Solar differential rotation lifts this degeneracy, leading to frequency splitting. Frequency splitting is conventionally represented as a polynomial expansion in the azimuthal order, $m$ of the modes. The odd-order coefficients in this expansion—denoted $a_1,\;a_2,\;a_3,\ldots$ and commonly known as the a-coefficients—encode information about solar differential rotation. The relationship between the $a$-coefficients and differential rotation can be formulated as an integral equation (see Equation~\ref{eq:a_int}). By determining the odd-order $a$-coefficients and solving this integral equation through an inversion procedure, the solar differential rotation can be inferred. In this study, we employ the Regularized Least Squares (RLS) method to perform the inversion and obtain the solar angular velocity $\Omega$. 
\par
\begin{figure}
\centering
   \includegraphics[scale=0.4]{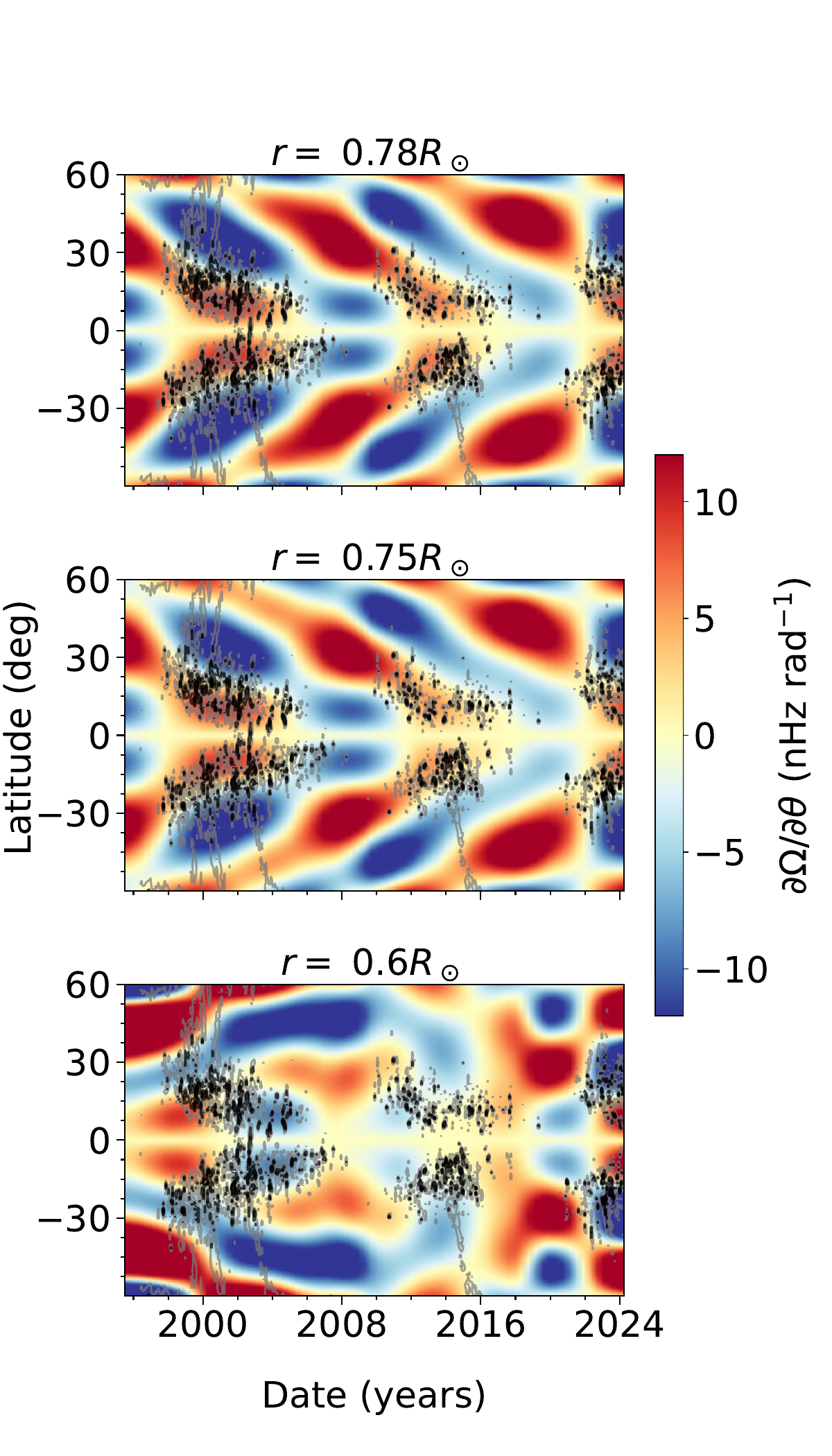}\hspace{2em}\includegraphics[scale=0.4]{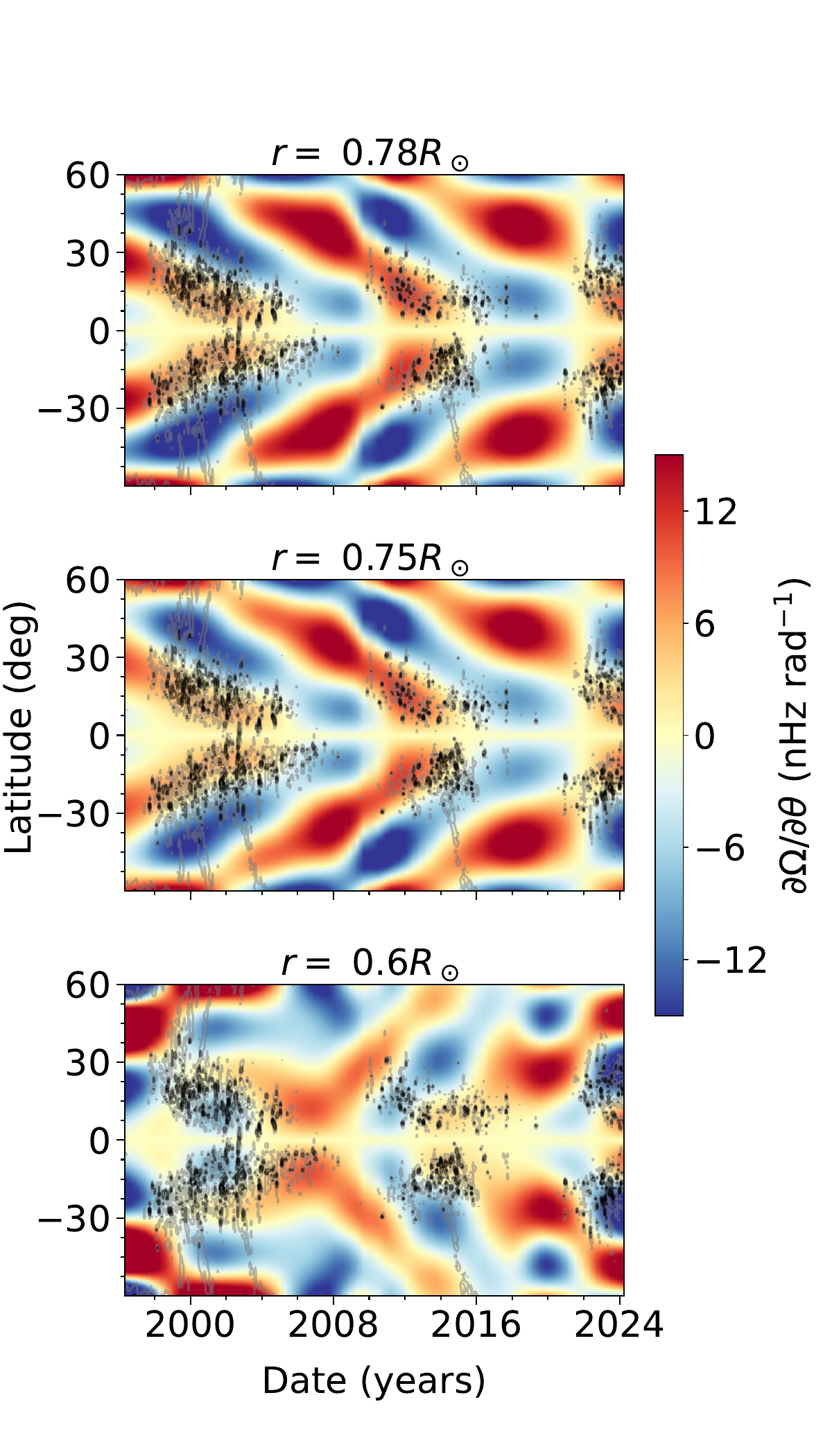}
    \caption{ We show $\partial\Omega/\partial\theta$, with Southern Hemisphere values multiplied by $-1$ to enforce hemispheric symmetry, as it naturally reverses sign across the equator. The analysis is based on $4 \times 72$-day GONG datasets (left panel) and combined MDI and HMI observations (right panel). The results are shown at three depths, $0.78$, $0.75$, and $0.6\ R_\odot$. At $0.78\,R_\odot$ and $0.75\,R_\odot$, the error in $\partial\Omega/\partial\theta$ is $1.2\,\mathrm{nHz\,rad^{-1}}$, rising to $3\,\mathrm{nHz\,rad^{-1}}$ at $0.6\,R_\odot$.
    To enhance the visibility of latitudinal patterns, the results are multiplied by $\sin\theta$. Magnetic-field contours for the same period are overlaid to highlight the correlation between the evolution of the toroidal magnetic field and $\partial\Omega/\partial\theta$.}
    \label{fig:dln_domega_all}
\end{figure}
\begin{figure}
    \includegraphics[scale=0.4]{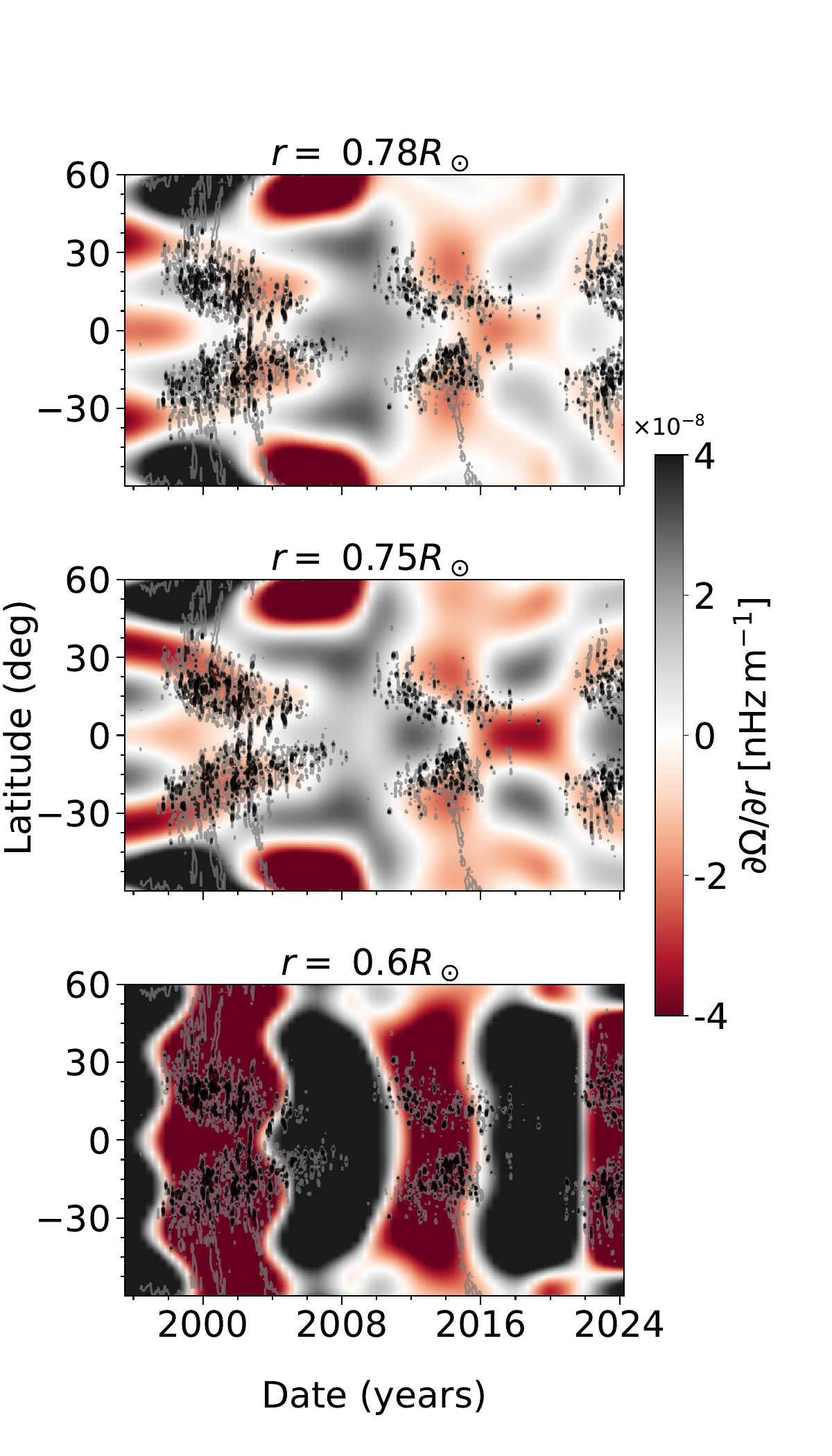} \hspace{2em}\includegraphics[scale=0.4]{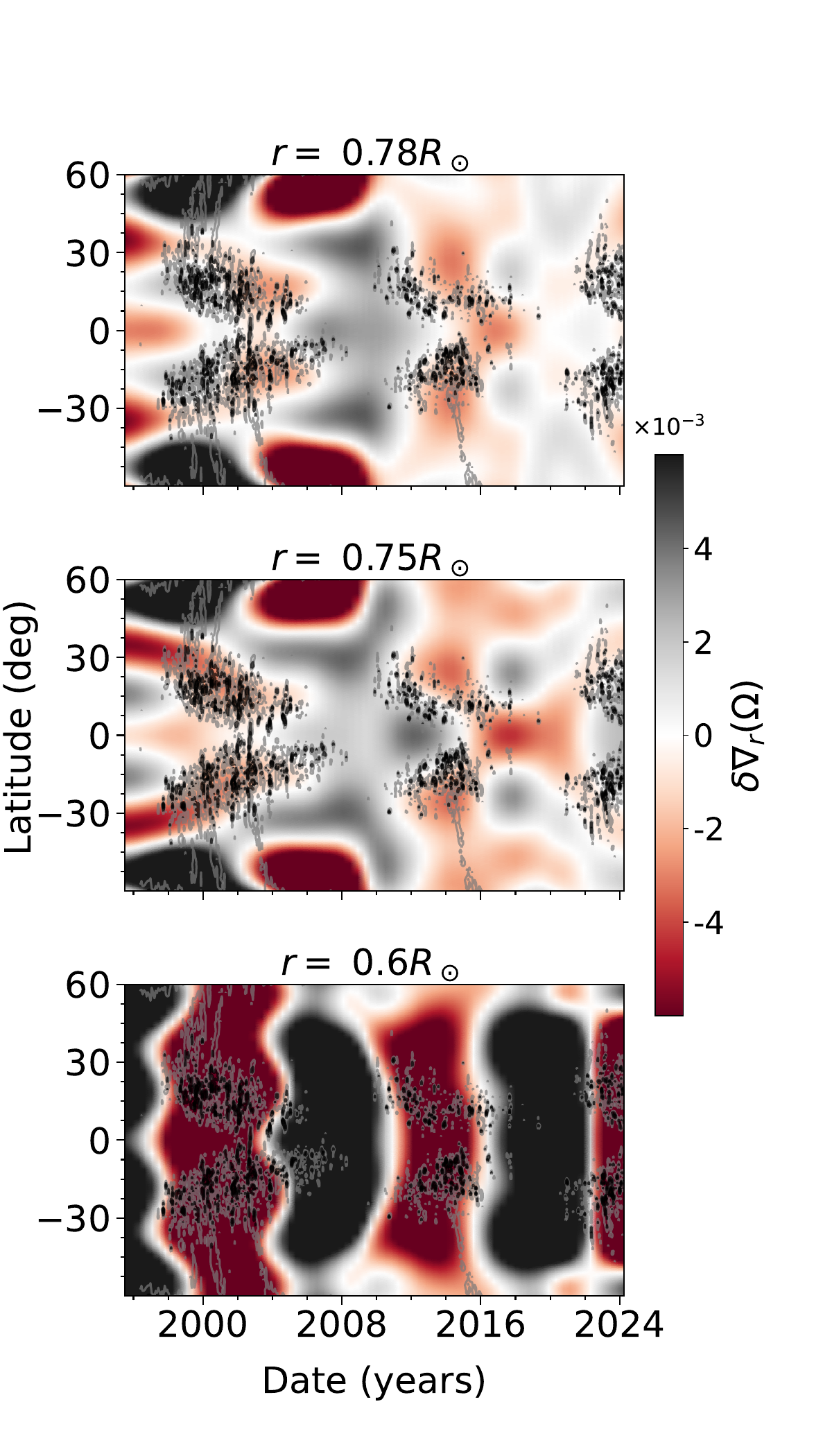}
    \caption{The left panel shows $\partial\Omega/\partial r$ as a function of the solar cycle. The right panel presents the dimensionless radial gradient, $\nabla_r\Omega$, at three depths $0.78$, $0.75$, and $0.6 R_\odot$, obtained from GONG $4 \times 72$-day data. The uncertainty in $\partial\Omega/\partial r$ is $6\times10^{-9}$ at depths of $0.78\,R_\odot$ and $0.75\,R_\odot$, increasing to $1.5\times10^{-8}$ at $0.6R_\odot$.}
    \label{fig:MDI+HMI}
\end{figure}

We use frequency splitting data from three instruments, SOHO Michelson Doppler Imager (MDI) \cite{scherrer95}, SDO Helioseismic Magnetic
Imager (HMI) \cite{hmi}, and Global Oscillation Network Group (GONG)   \cite{Harvey1996}. The GONG dataset \cite{GONG} spans from July 7, 1995, to April 28, 2024; MDI covers May 1, 1996, to July 11, 2010; and HMI extends from July 11, 2010, to April 28, 2024. These time ranges are sufficient to cover Solar Cycles 23 and 24 as well as the first half of Cycle 25 as it approaches its maximum. This long dataset, spanning more than two and a half solar cycles, is crucial for studying how differential rotation evolves with the solar cycle and for identifying possible correlations with the magnetic field measured on the surface. To improve the signal-to-noise ratio, frequency splittings are estimated from $4\times 72$-day segments available \cite{sylvain23}. The splittings derived from this time series are sufficiently precise to reduce noise, but the duration is still not long enough to fully suppress the time variations near the tachocline. We compute the radial gradient of the angular velocity, $\partial \Omega/\partial r$, the latitudinal gradient, $\partial \Omega/\partial \theta$, and the dimensionless radial gradient, $\nabla_{r}(\Omega)$, separately for Solar Cycles 23, 24, and 25. For each quantity, we subtract the corresponding cycle-averaged mean from its time series in order to emphasize its temporal variations over the solar cycle. Figure~\ref{fig:dln_domega_all} shows the variation of $\partial \Omega/\partial \theta$ over the solar cycle, based on the analysis of GONG data and the combined MDI and HMI datasets. The agreement between the two datasets is evident, with MDI+HMI and GONG yielding reasonably consistent results. Since $\partial \Omega/\partial \theta$ changes sign across the equator, its values in the Northern and Southern Hemispheres have opposite signs. To avoid confusion arising from this sign change, we multiply the Southern Hemisphere values of $\partial \Omega/\partial \theta$ by $-1$, ensuring that the values in both hemispheres are consistent and that hemispheric symmetry is preserved. Additionally, we present the variation of the dimensionless radial gradient, $\nabla_{r}(\Omega)$, and the radial gradient, $\partial \Omega/\partial r$, in the left and right panels of Figure~\ref{fig:MDI+HMI}, respectively, using the GONG dataset. In all of these plots, we overlay surface magnetic field contours to highlight the relationship between the observed magnetic field and variations in the gradient of rotation near the tachocline. Our results indicate that the equatorward migration of the rotational gradient at depths of $0.78$ and $0.75,R_\odot$ closely follows the surface migration of the solar magnetic field. However, at a depth of $\sim 0.6,R_\odot$, within the radiative interior, no clear correlation with the magnetic butterfly diagram is observed.

\begin{figure}

   \includegraphics[scale=0.4]{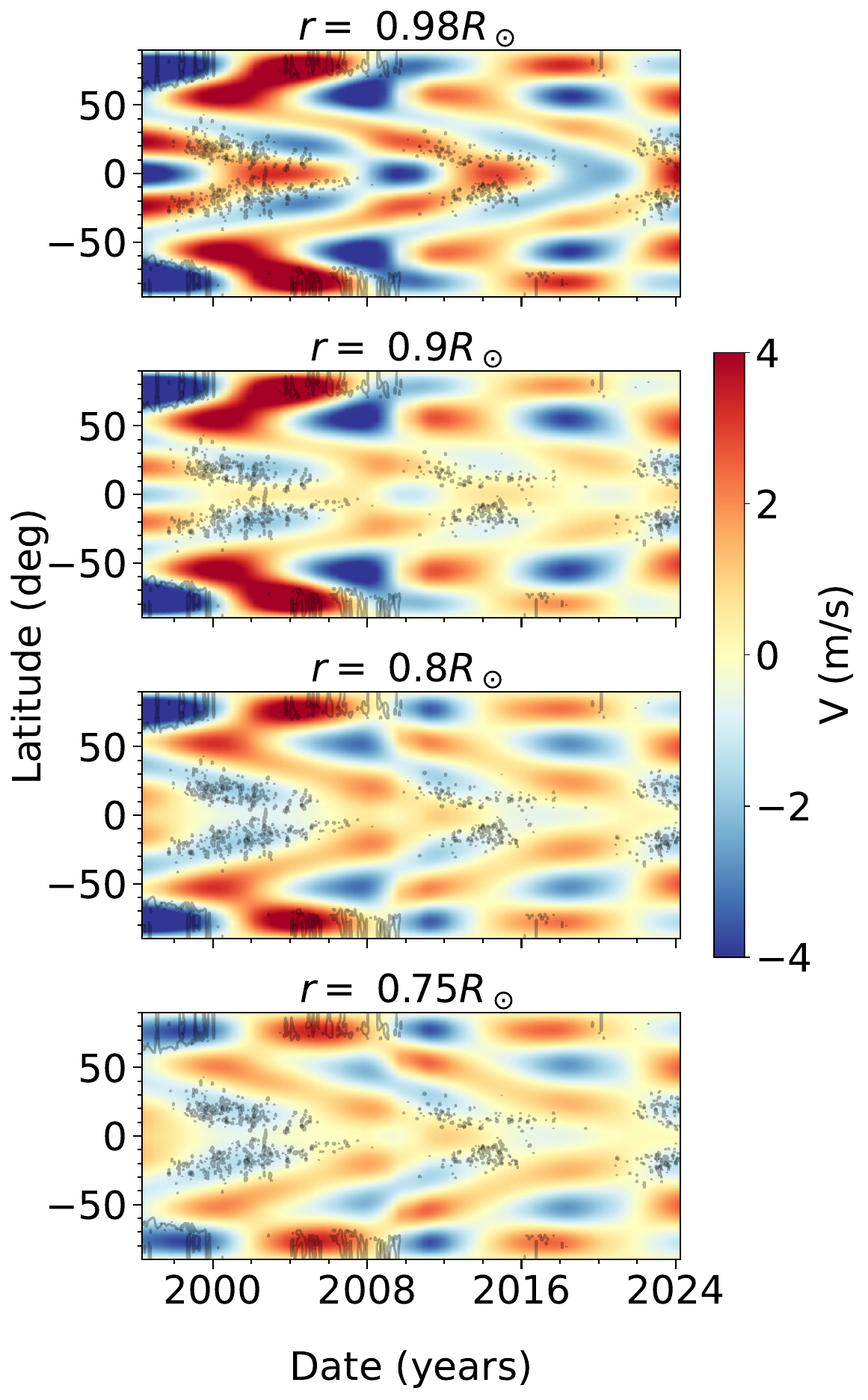}\hspace{5em}\includegraphics[scale=0.39]{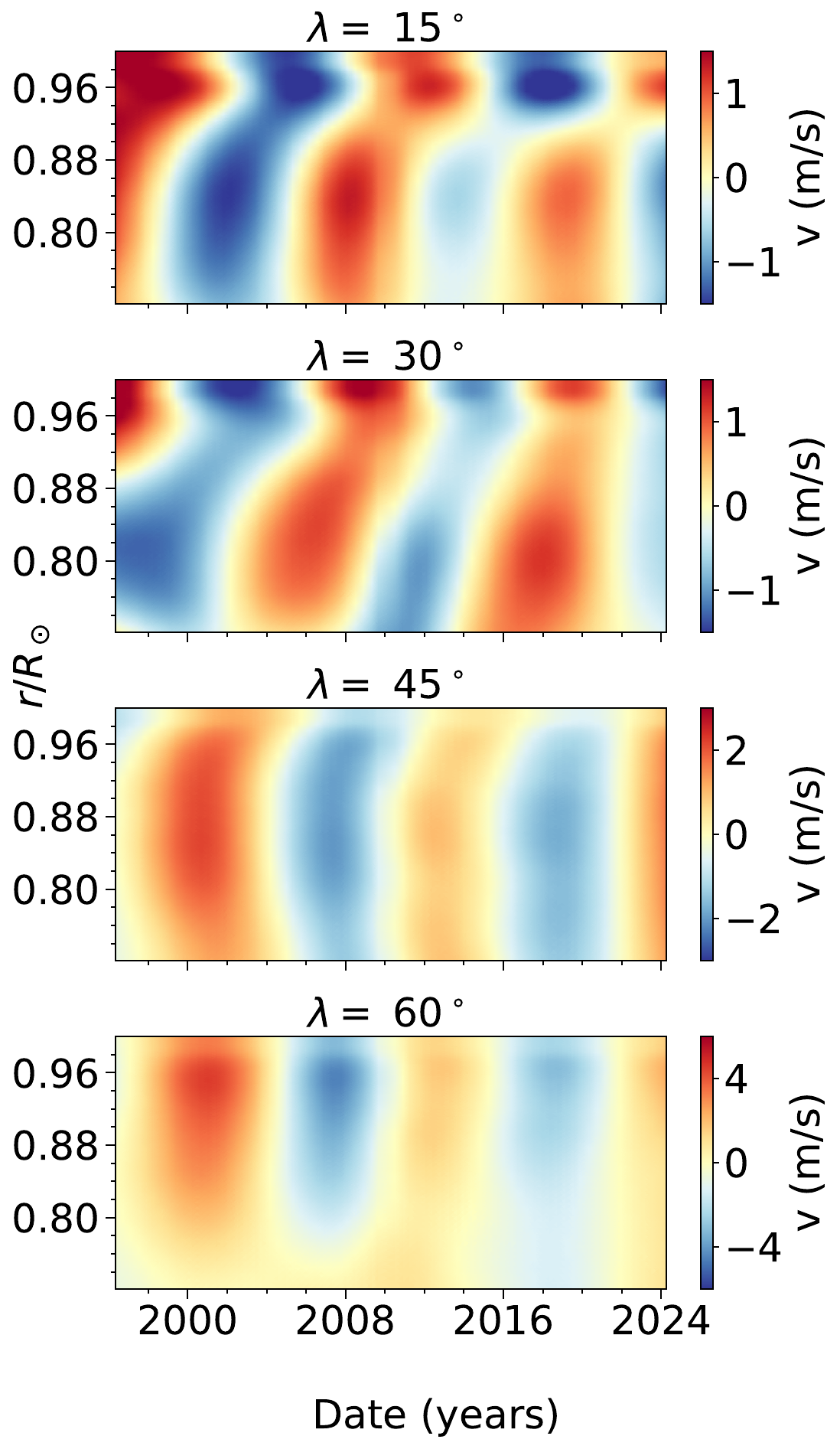}\\
   \includegraphics[scale=0.45]{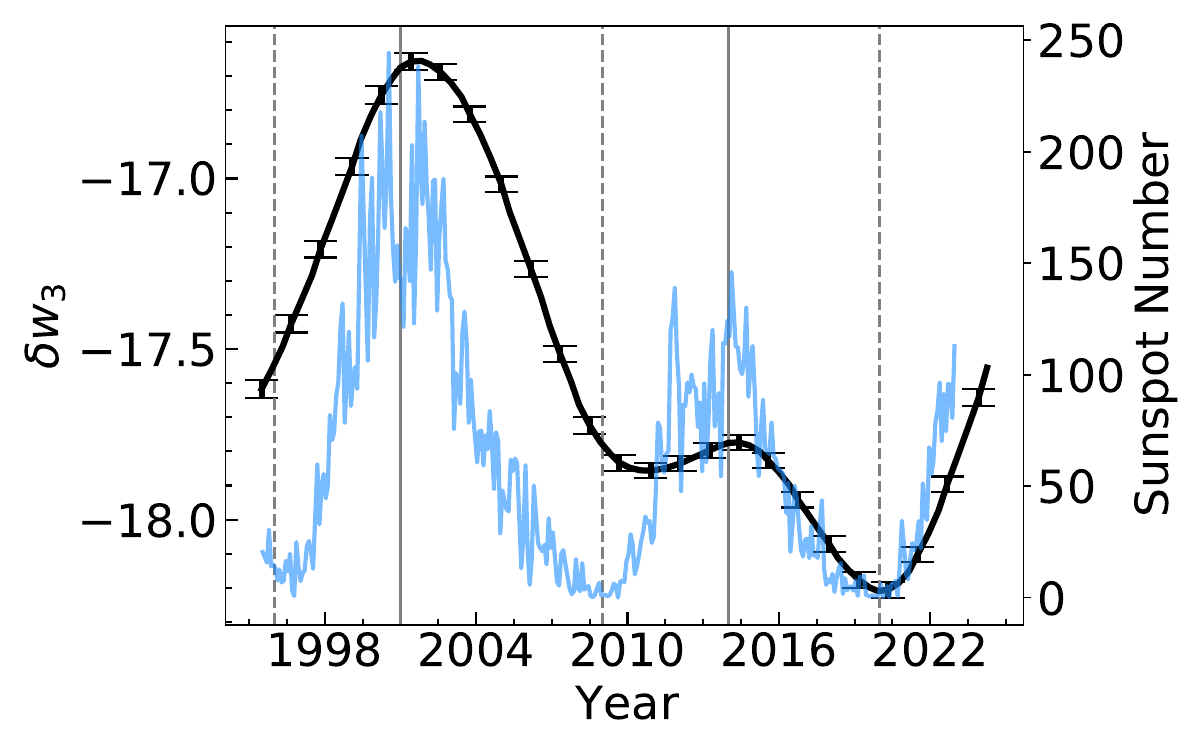}\hspace{2em}\includegraphics[scale=0.5]{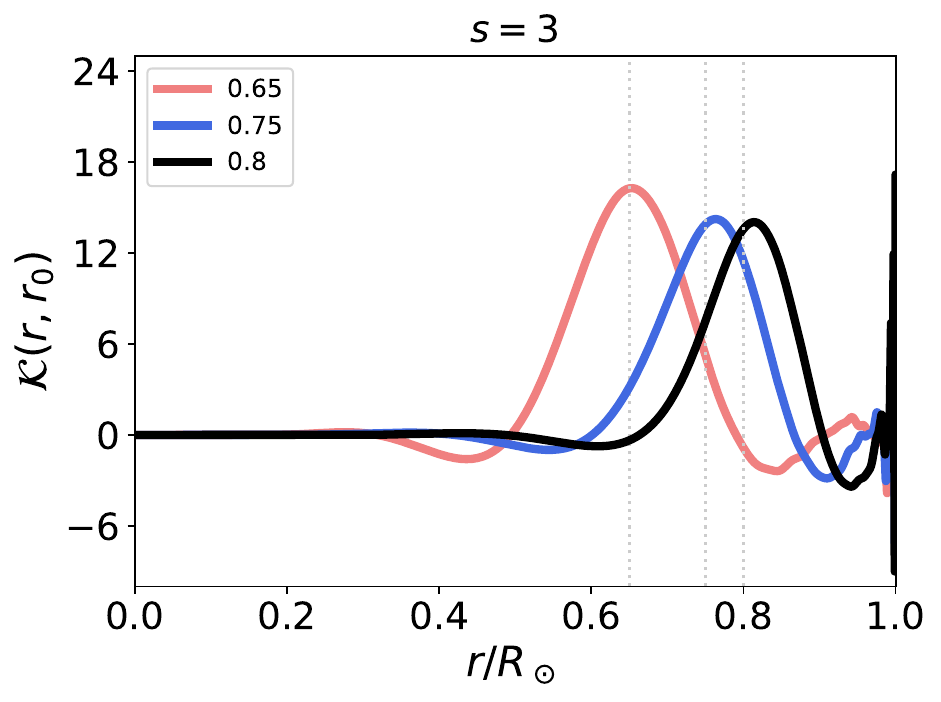}
    \caption{Upper left panel: zonal flows at several depths (indicated in each panel title), overlaid with the magnetic butterfly diagram to highlight the phase lag between flows at different depths and the magnetic pattern, using combined MDI and HMI datasets. Velocity measurements from the inversion have an accuracy of $0.4\,\mathrm{m\,s^{-1}}$ at $0.8\,R_\odot$, rising to $0.9\,\mathrm{m\,s^{-1}}$ at $0.6\,R_\odot$. Figure~14 of our previous work \cite{mandal_25} also presents error bars in the velocity measurement on a radius–latitude grid as a 2D plot.
    Top right panel: torsional oscillation variations with depth at selected latitudes (indicated in each panel title). At low latitudes, the signal emerges from the tachocline with a time lag before reaching the surface, whereas at high latitudes (near $50^\circ$) it appears first at the surface. This behavior is characteristic of a dynamo wave–like signature in solar torsional oscillations.
    Bottom left panel: we plot the jump in radial profile obtained from the inversion of $a_3$ coefficient between depths of $0.8\,R_\odot$ and $0.6\,R_\odot$.
    The dashed vertical line indicates the cycle minimum, while the solid vertical line marks the cycle maximum. The variation in sunspot number over the same period is shown by the solid light-blue line.
    Bottom right panel: Averaging kernels at selected depths (listed in the legend) are shown, with different colors representing different depths. }
    \label{fig:zonal_flow}
\end{figure}

We present the zonal flow in Figure~\ref{fig:zonal_flow}, overlaid with the sunspot number at several depths. At greater depths, the zonal flow shows little correlation with the magnetic field. In contrast, near the surface, the zonal flow is strongly correlated with the magnetic field, as shown in Figure~\ref{fig:zonal_flow}. Notably, there is a time lag between the equatorward migration of the positive branch of the zonal flow near the tachocline and the corresponding solar magnetic field observed at the surface. This behavior has been interpreted in previous studies \cite{sasha2019,mandal24_dw} as evidence of a dynamo-wave-like signature in the solar zonal flow: it originates at high latitudes near the tachocline, with one branch migrating equatorward and another poleward. This pattern is more clearly seen in the top-right panel of Figure~\ref{fig:zonal_flow}, which reveals a periodicity in the solar zonal flow throughout the entire convection zone. At low latitudes ($\leq 30^\circ$), perturbations originating near the tachocline take time to propagate to the surface, whereas at high latitudes ($\geq 45^\circ$) they reach the surface much more rapidly. While dynamo waves are typically observed in the evolution of the solar magnetic field \cite{parker_55_dynamo,yoshimura_1975}, these results suggest that a similar signature may also be present in the zonal flow. 
\par
\begin{figure}
    \centering
    \includegraphics[scale=0.4]{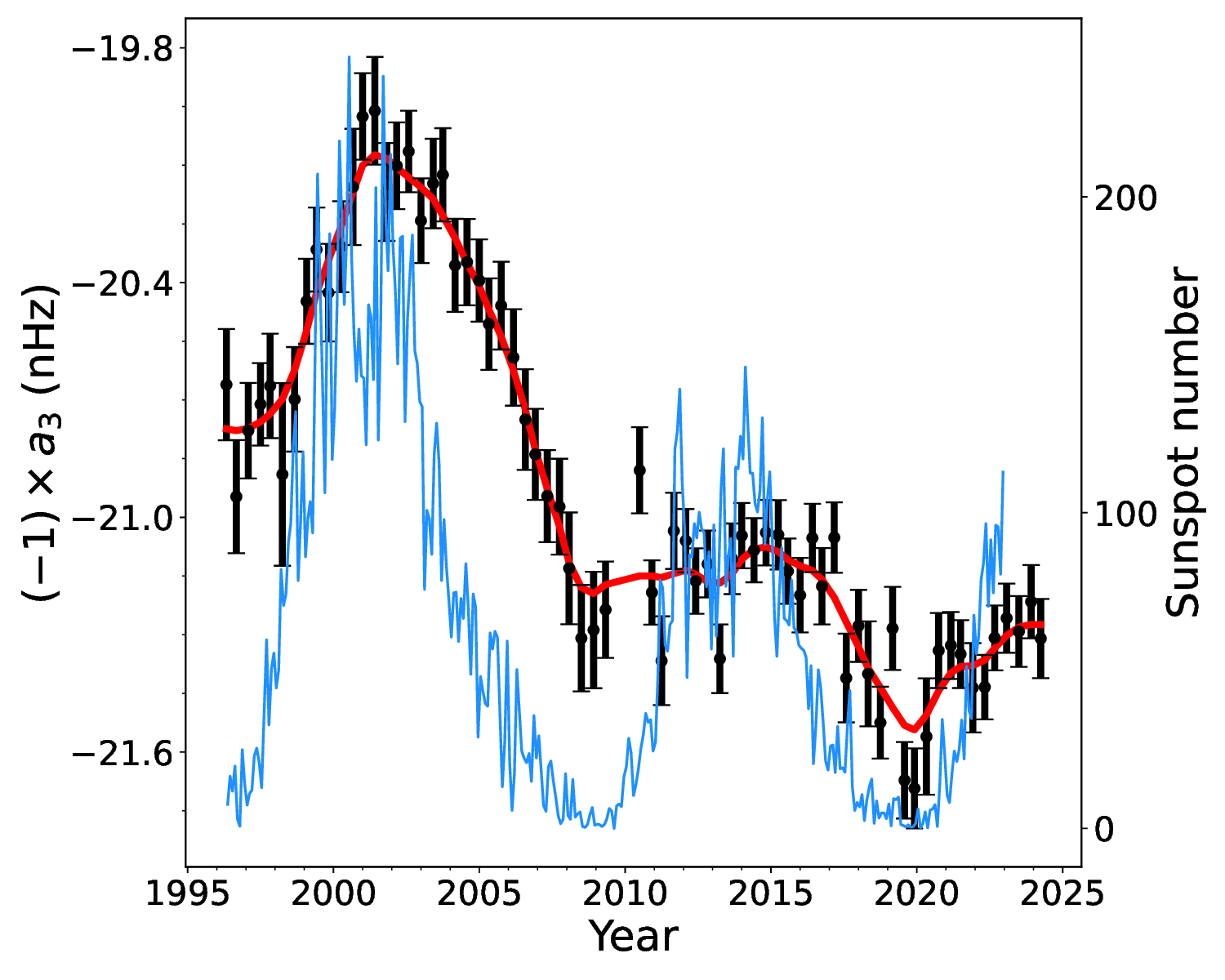}
    \caption{We calculate the average $a$-coefficients (shown by black markers, with black error bars) for modes with turning points between $0.7$ and $0.8,R_\odot$ and plot them alongside the sunspot number (light blue line). The variation of the $a$-coefficients is further smoothed with a Gaussian filter and shown as a red solid line. The plot is based on the combined MDI and HMI $4\times 72$-day datasets.}
    \label{fig:a3_var}
\end{figure}

 If the tachocline is responsible for generating and storing the solar toroidal magnetic field, we expect variations in the properties of the tachocline with the solar cycle. As noted in previous studies \cite{basu_2024}, a jump in the rotation rate near the tachocline can be inferred by fitting the $a_3$ coefficient, which encapsulates most of the information about the tachocline. Here, we directly invert for the radial profile corresponding to the $a_3$ coefficient. We then estimate the jump in its values between $0.8R_\odot$ and $0.6 R_\odot$ of that radial profile where, $R_\odot$ represents the solar radius, and present the results in Figure~\ref{fig:zonal_flow}. We find that radial profile from the inversion of $a_3$ exhibits significant solar-cycle–dependent variations in the tachocline region: the jump is more pronounced during Cycle 23, weaker in Cycle 24, and appears to be strengthening again in Cycle 25 based on initial data as the cycle approaches its maximum. This jump near the tachocline closely tracks the strength of the solar cycle. To verify this idea instead of inversion we consider signal in the $a_3$-coefficient itself. We compare the $a_3$ coefficients for modes with turning points between $0.8$ and $0.7,R_\odot$ and plot their variation over the solar cycle in Figure~\ref{fig:a3_var}. Despite some scatter, smoothing reveals a clear trend that agrees with the inversion results in Figure~\ref{fig:zonal_flow}. The results in Figure~\ref{fig:a3_var} show a variation similar to that obtained from the inversion using all $a$-coefficients for the radial profile in Figure~\ref{fig:zonal_flow}.
If the two are correlated, this relationship could potentially serve as a predictor of the next cycle’s amplitude. However, because the variations occur almost simultaneously, the predictive capability is limited. Whether the correspondence between the $a_3$ jump and solar cycle strength reflects a physical connection or is merely coincidental remains uncertain, and further helioseismic observations over multiple cycles will be needed to resolve this. At high latitudes, our results are subject to greater uncertainty, primarily due to the limited resolution determined by the width of the averaging kernel. The resolution decreases with depth, since only a small number of p-modes penetrate into the deeper layers (see the right panel of Figure~\ref{fig:zonal_flow}). The width of the averaging kernel sets the effective resolving power of our analysis: as shown in the bottom-right panel of Figure~\ref{fig:zonal_flow}, the kernel width increases with depth, leading to poorer resolution—a general limitation inherent to traditional helioseismic analyses.
 

\section*{Discussion}
The solar magnetic field cannot be measured directly in the solar interior and can only be inferred indirectly. While the emergence of sunspots on the solar surface may appear chaotic, the underlying magnetic field is highly organized, evolving coherently over large spatial scales and on the characteristic timescale of the 11-year solar cycle. Similarly, solar torsional oscillations vary coherently on the 11-year cycle and exhibit a strong correlation with the latitudinal migration of sunspots. This observed correspondence between torsional oscillations and surface magnetic activity suggests that zonal flows and magnetic fields are dynamically coupled through Lorentz-force feedback. The solar dynamo represents the continuous conversion of mechanical energy into magnetic energy; since Ohmic dissipation steadily depletes magnetic fields, mechanical energy must work against the Lorentz force to maintain the dynamo. Although direct measurements of magnetic fields in the solar interior are not possible, helioseismology enables the detection of torsional oscillations deep within the Sun. Analysis of deep torsional oscillation has revealed a dynamo-wave–like signature in the zonal flow (top panel of Figure~\ref{fig:zonal_flow})\cite{sasha2019,mandal24_dw}. This provides one of the first indications that the solar dynamo may have a deep-seated origin. Such dynamo waves were first proposed in theoretical models of solar dynamo \cite{parker_55_dynamo,yoshimura_1975}.  

We find a strong and statistically significant correlation between variations in the gradient of rotation near the tachocline and the solar magnetic butterfly diagram observed at the surface. Although torsional oscillations near the tachocline exhibit a slight time lag relative to the butterfly diagram—a feature consistent with the dynamo-wave–like signature discussed earlier (Figure~\ref{fig:zonal_flow})—both the radial and latitudinal gradients remain strongly correlated with it. We find that the butterfly diagram is associated with a positive dimensionless radial gradient between depth of $0.97R_\odot$ and $0.92R_\odot$ \cite{mandal_25}, but with negative values near the tachocline (Figure~\ref{fig:dln_domega_all}). This indicates that the processes responsible for torsional oscillations at the surface also operate in deeper layers, supporting the view that the solar dynamo extends throughout the convection zone. For the first time, we demonstrate that both the radial and latitudinal gradients display torsional-oscillation–like patterns in the vicinity of the tachocline, further reinforcing that the dynamo is not confined to the near-surface shear layer but permeates the entire convection zone. There is ongoing debate about whether the tachocline plays a central role in the solar dynamo. If the Sun’s cyclic magnetic field truly originates in the tachocline, one would expect cyclic modulations of the local angular velocity, similar to those observed at the surface \cite{howard80}. In this work, we demonstrate precisely that. 
\par
Previous work \cite{vasil2024} suggested that magneto-rotational instability (MRI) may be responsible, proposing that the near-surface shear layer (NSSL) could serve as the seed region for the solar dynamo instead of the tachocline. However, the foundation of their argument—that helioseismology shows torsional oscillations exist only within the outer $5$–$10\%$ of the solar radius—is not supported by our observations. As shown in Figure~\ref{fig:zonal_flow}, torsional oscillations extend throughout the entire convection zone. Thus, their conclusion that the NSSL is the seed of the dynamo, based on MRI simulations that are assumed to match torsional oscillations confined to the NSSL, may not be valid in light of our results. Furthermore, their analysis identifies MRI as the dominant mechanism driving the solar dynamo by filtering out large-scale baroclinic effects, small-scale convection, and nonlinear dynamo feedback—an assumption that may not be valid in light of our findings. This concern is reinforced by a recent study \cite{brandenburg_2025}, which reported that turbulent magnetic diffusivity may be too high for MRI to be excited in NSSL.

As discussed earlier, information about the tachocline is primarily contained in the $a_3$ coefficient of frequency splitting. While most other $a$-coefficients vary smoothly across the tachocline, $a_3$ exhibits a pronounced variation, making it the most sensitive diagnostic of this region. Using our 1.5D RLS inversion method, we analyze the rotation profile inferred from the $a_3$ coefficients and find that the jump in rotation across the tachocline follows the solar cycle. Specifically, the jump was strongest during Solar Cycle 23, weaker in Cycle 24, and has strengthened again in Cycle 25 as it approaches maximum activity. This trend closely mirrors the variation in sunspot numbers, suggesting a direct link between tachocline dynamics and the strength of solar activity. Although the variations in $a_3$ and the solar cycle indices occur nearly simultaneously, limiting predictive capability, these results provide valuable insight into the operation of the solar dynamo. Although our results do not pinpoint the specific dynamo mechanism operating in the Sun, they underscore the pivotal role of the tachocline. The strong rotational shear in this region likely facilitates the amplification of poloidal fields into toroidal fields. Since Ohmic dissipation continuously depletes magnetic fields, a persistent conversion of mechanical energy into magnetic energy—driven by flows working against the Lorentz force—is required to sustain dynamo action. In this work, we analyze only the odd-order $a$-coefficients to investigate variations in solar rotation. The even-order splitting coefficients ($a_2,,a_4,,\ldots$) can be studied in a similar manner over the solar cycle, which would allow us to probe changes in the solar sound speed \cite{antia_2001,baldner_2008}. Inverting these coefficients makes it possible to infer the solar-cycle variations of sound speed anomalies. With an appropriate choice of equation of state, the corresponding variations in temperature and entropy can also be estimated. Similar studies have been carried out previously \cite{brun_2010}. Analyzing both the odd and even $a$-coefficients provides a more complete picture of how rotation, temperature, and entropy vary over the solar cycle. These observational inferences can be directly compared with MHD simulations\cite{pipin2019}, allowing us to assess how well different dynamo models reproduce the solar cycle and, in turn, to better constrain the dynamo operating in the Sun.

\section*{Methods}
 The availability of rotational p-mode frequency data over various time lengths \cite{sylvain23} allows us to study the tachocline region of the Sun with a great accuracy. Specifically, we use frequency fitting results from time series of $4\times 72-$days based on HMI, MDI, and GONG observations. These datasets are accessible at the Joint Science Operations Center (JSOC) under the following data series names: 1) HMI: \texttt{su\_sylvain.hmi\_V\_sht\_modes\_asym\_v7}, 2) MDI: \texttt{su\_sylvain.mdi\_V\_sht\_modes\_asym\_v7}, 3) GONG: \texttt{su\_sylvain.gong\_V\_sht\_modes\_asym\_v7}. The GONG dataset spans from May 7, 1995, to March 23, 2024; MDI covers May 1, 1996, to March 20, 2011; and HMI extends from April 30, 2010, to August 14, 2024. There is approximately a one-year overlap between the MDI and HMI datasets, during which we prioritize HMI data. The frequency of a solar acoustic mode is typically denoted as $\nu_{n\ell m}$, where $n$, $\ell$, and $m$ correspond to the radial order, harmonic degree, and azimuthal order, respectively. In a spherically symmetric Sun, these frequencies are degenerate in $m$. However, solar rotation breaks this degeneracy, and the resulting frequency shifts can be expressed as:
\begin{equation}
    \omega_{n\ell m}/(2\pi)= \nu_{n\ell}+\sum_{s=1,3,5,\dots,s_\text{max}} a_{s}(n,\ell)\mathcal{P}_{s}^{\ell}(m)
\end{equation}
where $\nu_{n\ell}$ is the central frequency of the multiplet, $a_s(n, \ell)$ are the rotational splitting coefficients, and $\mathcal{P}_{s}^{\ell}(m)$ are the polynomials of degree $s$, and the recursion relation of these polynomials can be found in previous work \cite{schou_94_inv}. Inversions based on global modes are sensitive only to the equatorially symmetric component of rotation; therefore, any hemispheric asymmetry in the rotation profile may not be captured in our analysis. Before analyzing the observed frequency-splitting data, we validated our methodology through forward modeling and inversion of synthetic splittings generated from the rotation profile of a dynamo model \cite{pipin2019}, as detailed in our previous studies \cite{mandal_25,mandal24_dw}. Although various inversion techniques exist \cite{antia_1998,schou98}, we employ the 1.5D RLS method in the present study. A brief overview of our analysis procedure is provided below. The a-coefficients are related to differential rotation as follows:
\begin{equation}
    a_{s}(n,\ell)=\int_{0}^{R_\odot} K_{n,\ell,s}(r)w_{s}(r)r^{2}dr,
    \label{eq:a_int}
\end{equation}
where $s$ is harmonic degree and $w_s(r)$ is the radial profile used to represent the rotation profile as
\begin{equation}
    v(r,\theta)=\sum_{s=1,3,5,\ldots}w_s(r)\partial_{\theta}Y_{s0}(\theta,\phi),
    \label{eq:vel}
\end{equation}
The latitudinal dependence of the rotational velocity $v(r,\theta)$ is captured through the derivative of the spherical harmonic function $\partial_{\theta}Y_{s0}(\theta,\phi)$.
$K_{s}(n,\ell)$ is the sensitivity kernel, given by the following expression (its derivation is provided in the appendix of our previous work \cite{mandal24_dw}).
\begin{equation}
    K_s(n,\ell;r)=4\ell(\ell+1)(2\ell+1)\frac{2s+1}{4\pi}\frac{\{(2\ell-1)!\}^2}{(2\ell-s)!(2\ell+s+1)!}\mathcal{P}_s^{\ell}(1)\rho\left[ U^2-2UV+\frac{2\ell(\ell+1)-s(s+1)}{2}V^2\right]
    \label{eq:kernel_expr}
\end{equation}
 where $U(r)$ and $V(r)$ are the radial and horizontal components of the displacement vector, $\boldsymbol{\xi}_{k}$. where $k$ corresponds to the multi-index $(n,\ell,m)$ in a spherically symmetric Sun, as follows
\begin{equation}
\boldsymbol{\xi}_{k}=U(r)Y_{\ell m}\hat{r}+V(r)\boldsymbol{\nabla}_{h}Y_{\ell m},
\label{eq:eigFn}
\end{equation}
 where $\boldsymbol{\nabla_h}$ is the horizontal derivative, $\boldsymbol{\nabla_h=\hat{\theta}\partial_\theta+\frac{\hat{\phi}}{\sin\theta}\partial_\phi}$. We minimize the following misfit function to obtain $w_s(r)$ 
\begin{equation}
\chi^2=\sum_{n,\ell}\frac{1}{\sigma_{n,\ell}^{2}}\left(a_{n,\ell,s}-\int K_{n,\ell,s}(r)w_{s}(r)r^{2}dr\right)^{2}+ \mathrm{Regularization},
\label{eq:chi_app}
\end{equation}
The regularization term can be chosen to enforce either first- or second-derivative smoothing. We find that first-derivative smoothing performs better in the deep convection zone, where the sensitivity of the data is lower. In contrast, second-derivative smoothing tends to propagate gradients from high-sensitivity regions into low-sensitivity regions, which can introduce artifacts \cite{antia_1998}. This issue is addressed by first-derivative smoothing; therefore, we employ first-derivative regularization in this work.  
\begin{equation}
    \mathrm{Regularization}=\lambda\int\left(\frac{d w_{s}}{d r}\right)^{2}r^{2}dr
\end{equation}
We express $w_{s}$ in terms of the B-splines in radius as follows
\begin{equation}
 w_{s}(r)=\sum_{i=1}^{N}b_{i}B_{i}(r),\label{eq:ws_Bspline}
\end{equation}
$N$ corresponds to the total number of unknowns. If we substitute Equation \ref{eq:ws_Bspline} into Equation \ref{eq:chi_app}, we get
\begin{align}
\chi^2=\sum_{n,\ell}\frac{1}{\sigma_{n,\ell}^{2}}\left(a_{s}(n,\ell)-\sum_{i}\int K_{n,\ell,s}(r)b_{i}B_{i}(r)r^{2}dr\right)^{2}+\lambda\int\left(\frac{d\sum_{i}b_{i}B_{i}(r)}{d r}\right)^{2}r^{2}dr
\end{align}
We minimize the above equation with respect to the unknown coefficients $b_j$, resulting in a set of linear equations that can be expressed in matrix form. Solving this matrix equation yields the coefficients $b_j$. A detailed derivation of these matrix equations can be found in the appendix of our previous work \cite{mandal24_dw}. The velocity profile can be obtained from Equation~\ref{eq:vel} by substituting the determined $b_j$ coefficients. The rotation angular velocity $\Omega$ is then related to the velocity through
\begin{equation}
    \Omega(r,\theta)=V(r,\theta)/(r\sin\theta).
\end{equation}

\bibliography{reference}



\section*{Funding}
This work was partially supported by NASA grants 80NSSC20K1320, 80NSSC20K0602, and 80NSSC22M0162. K.M. acknowledges support from NASA grant 80NSSC25K7755.

\section*{Author contributions statement}
K.M. and A.K. have conceived the ideas, K.M. analyzed the data, and all the authors contributed to writing the manuscript.

\section*{Additional information}
The authors declare that they have no financial or commercial conflicts of interest.


\end{document}